\documentclass[aps,prl,twocolumn,superscriptaddress,showpacs]{revtex4}

\usepackage{CJK}

\usepackage{graphicx}

\usepackage{dcolumn}

\usepackage{bm}
\usepackage[normalem]{ulem}
\usepackage{color}

\usepackage{epstopdf}
\newcommand{\bup}{{\bu_{_\perp}}}

\newcommand{\ii}{{\rm i}}
\newcommand{\bx}{\mathbf{x}}

\newcommand{\bq}{\mathbf{q}}
\newcommand{\bv}{\mathbf{v}}

\newcommand{\br}{\mathbf{r}}

\newcommand{\bu}{\mathbf{u}}

\newcommand{\sep}{ \ \ \ , \ \ \ }

\newcommand{\beq}{\begin{equation}}
\newcommand{\eeq}{\end{equation}}
\newcommand{\beqn}{\begin{eqnarray}}
\newcommand{\eeqn}{\end{eqnarray}}
\newcommand{\pp}{\partial}
\newcommand{\dd}{{\rm d}}

\newcommand{\eq}{Eq.\ }

\newcommand{\eqs}{Eqs }
\newcommand{\fig}{Fig.\ }

\newcommand{\dx}{\partial_x}
\newcommand{\dt}{\partial_t}

\newcommand{\la}{\langle}

\newcommand{\ra}{\rangle}

\begin{document}

\begin{CJK*}{GBK}{}
	
	\title{Incompressible polar active fluids in the moving phase
	}
	
	\author{Leiming Chen}
	\email{leiming@cumt.edu.cn}
	\affiliation{School of Physical Science and Technology, China University of Mining and Technology, Xuzhou Jiangsu, 221116, P. R. China}
	\author{Chiu Fan Lee}
	\email{c.lee@imperial.ac.uk}
	\affiliation{Department of Bioengineering, Imperial College London, South Kensington Campus, London SW7 2AZ, U.K.}
	\author{John Toner}
	\email{jjt@uoregon.edu}
	\affiliation{Department of Physics and Institute of Theoretical
		Science, University of Oregon, Eugene, OR $97403$}
	
	\begin{abstract}
	We study universal behavior in the moving phase of a generic system of motile particles with alignment interactions in the incompressible limit for spatial dimensions $d>2$. Using a dynamical renormalization group analysis, we obtain the exact dynamic, roughness, and anisotropy exponents that describe the scaling behavior of such incompressible systems. 
This is the first time a compelling argument has been given for the exact values of the anomalous scaling exponents of a flock moving through an isotropic medium in $d>2$.
	\end{abstract}
	\maketitle
\end{CJK*}

Unshackled
by the fluctuation-dissipation relation, non-equilibrium systems exhibit many novel behaviors impossible in equilibrium systems. One of the most striking examples is the existence of long-ranged order associated with a broken continuous symmetry in two dimensions (2D) -- a phenomenon forbidden in equilibrium systems by the Mermin-Wagner theorem \cite{MW}. Collective motion, or ``flocking'', can therefore exist in active matter, even in two dimensions \cite{TT1,TT2, TT3,TT4, NL}.  Active matter is currently receiving intense attention from both physics and biology communities due to its importance to non-equilibrium physics, developmental biology, cell and tissue mechanics, and ecology \cite{rmp}. However, the study of universal behavior of active matter is plagued by the emergence of nonlinearities in the generic equation of motion (EOM).
In the case of polar active fluids, a crucial simplifying assumption that enables analytical progress is the incompressibility condition \cite{us_njp,us_NC}.
Incompressibility is {\it not} merely a theoretical contrivance; not only can it be readily simulated \cite{ramaswamy_jcomputphys15},  it can arise in a variety of real experimental situations, such as 1) systems  with strong repulsive short-ranged interactions between the active particles.  Incompressibility has, in fact, been assumed in, e.g.,   recent experimental studies on cell motility \cite{wensink_pnas12}; and 2) systems with long-ranged repulsive interactions; here, true incompressibility is possible. Long-ranged interactions are quite reasonable in certain contexts: birds, for example, can often see all the way across a flock \cite{turner}.

By eliminating the density fluctuations, we have recently shown that generic polar active fluids can exhibit a continuous order-disorder phase transition with the corresponding critical behavior belonging to a novel universality class \cite{us_njp}; and in the ordered phase in 2D, the static (equal-time) behavior of the system can be mapped onto the Kardar-Parisi-Zhang surface growth model  \cite{us_NC}. Here, we use dynamical renormalization group analysis to elucidate the universal behavior of generic incompressible polar active fluids in the ordered phase in spatial dimensions $d>2$, and reveal the surprising connection between infinitely compressible (Malthusian), compressible, and incompressible polar active fluids.

Specifically we find the following scaling law of the velocity correlation functions for incompressible polar active fluids at spatial dimensions $d>2$:
\begin{eqnarray}
&&\langle\bu_{_\perp}(\mathbf 0,0)\cdot\bu_{_\perp}(\br,t)\rangle\nonumber\\
&\sim&\left\{
\begin{array}{ll}
r_{_\perp}^{2\chi},&|x-v_1t|\ll r_{_\perp}^{\zeta}, |t|\ll r_{_\perp}^z\\
|x-v_1t|^{2\chi\over\zeta},&|x-v_1t|\gg r_{_\perp}^{\zeta}, |t|^{1\over 2}\\
|t|^{2\chi\over z},&|t|\gg r_{_\perp}^z, |t|\gg |x-v_1t|^2
\label{uscale}
\end{array}\,,
\right.
\end{eqnarray}
where the characteristic speed $v_1$ is a phenomenological, system-dependent  parameter, $\hat{\bf x}$ is along the mean velocity of the system, ``$\perp$'' denotes components perpendicular to $\hat{\bf x}$, $t$ is time, and $\bu$ is the coarse grained velocity. We have determined the {\it exact} values of the scaling exponents:
\beq
\zeta={d+1\over 5}\sep z={2(d+1)\over 5} \sep \chi={3-2d\over 5} \ .
\label{exact}
\eeq
Identical results have been conjectured for compressible polar active fluids \cite{TT1,TT5}  and Malthusian flocks \cite{TT5}, but only  in the latter case, and even there only in  2D, can  a compelling argument for them be made. The results we present here make isotropic incompressible flocks  the first polar active system with underlying isotropic symmetry in $d>2$ for which the exact scaling laws have been determined.

{\it Hydrodynamic model.}
We start with the hydrodynamic model for {\it compressible}
polar active fluids  without momentum conservation \cite{TT1, TT3, TT4, NL}:
\begin{widetext}\begin{eqnarray}\partial_{t}\bv+\lambda_1(\bv\cdot\nabla)\bv+\lambda_2(\nabla\cdot\bv)\bv+\lambda_3 \nabla(|\bv|^2)=U\bv -\nabla P -\bv\left( \bv \cdot \nabla P_2 \right) +\mu_{_B} \nabla(\nabla\cdot \bv)+ \mu_{_T}\nabla^{2}\bv +\mu_{2}(\bv\cdot\nabla)^{2}\bv+\mathbf{f}\nonumber \\\label{EOM}\end{eqnarray}\end{widetext}
\begin{eqnarray}
\partial_t\rho +\nabla\cdot(\bv\rho)=0
\label{conservation}
\end{eqnarray}
where $\bv(\br,t)$, and $\rho(\br,t)$ are respectively the coarse grained continuous velocity and density fields.
All of the parameters $\lambda_i (i = 1 \to 3)$,
$U$, the ``damping coefficients" $\mu_{B,T,2}$, the  ``isotropic pressure'' $P(\rho,
v)$ and the  ``anisotropic Pressure'' $P_2 (\rho, v)$
are  functions of the density $\rho$ and the
magnitude $v\equiv|\bv|$ of the local velocity.

The $U$ term makes the local
$\bv$ have a nonzero magnitude $v_0$
in the ordered phase, by 
having $U>0$ for $v<v_0$,
$U=0$ for $v=v_0$, and $U<0$ for $v>v_0$.
The $\mathbf{f}$ term is a random
driving force representing the noise. It is  assumed to be Gaussian with
white noise correlations:
\begin{eqnarray}
   \langle f_{i}(\br,t)f_{j}(\br',t')\rangle=2D
\delta_{ij}\delta^{d}(\br-\br')\delta(t-t')
\label{white noise}
\end{eqnarray}
where the ``noise strength" $D$ is a constant parameter of the system, and $i,j$ denote
Cartesian components.

We now take the incompressible limit by taking  {\it  the isotropic pressure $P$ only}
to be extremely sensitive to departures from the mean density $\rho_0$. 
Making $U(\rho, v)$ and $P_2(\rho, v)$
extremely sensitive to changes in $\rho$ as well proves to destabilize the system by generating a ``banding instability"\cite{unstable}, similar to the instability found in compressible active fluids around the onset of collective motion \cite{banding}.
Since we wish to focus on stable flocks, we will not consider this possibility further.

Focusing here on the case in which {\it only} the isotropic pressure $P$ becomes
extremely sensitive to changes in the density, we see that, in this limit, in which the
isotropic pressure will
suppress density fluctuations extremely effectively, changes in the density will be
too small to affect  $U(\rho, v)$, $\lambda_{1,2,3}(\rho, v)$, $\mu_{_B,_T,2}(\rho, v)$,
and $P_2(\rho, v)$.  As a result, 
all of them 
become functions only of the
speed $v$; their $\rho$-dependence will drop out since $\rho$ will be
essentially constant.

The suppression of density fluctuations by the isotropic pressure $P$ reduces the continuity equation (\ref{conservation})  
to the familiar condition 
for incompressible flow,
\begin{eqnarray}
\nabla\cdot\bv=0 \,,
\label{inc}
\end{eqnarray}
which can, as in simple fluid mechanics, be used to determine the isotropic pressure $P$.

The result of the above observations is the EOM:
\begin{eqnarray}
\partial_{t}\bv+\lambda(\bv\cdot\nabla)\bv=U(v)\bv -\nabla P   -\bv\left( \bv \cdot \nabla P_2 \right)\nonumber\\
+\mu_{_\perp}\nabla^2_{_\perp}\bv +\mu_x\partial_x^{2}\bv+\mathbf{f}\,,
\label{vEOM2}
\end{eqnarray}
where the statistics of the noise term are given by (\ref{white noise}), and the pressure $P$ is determined by the incompressibility condition (\ref{inc}). We've also defined $\lambda\equiv\lambda_1(|\bv|=v_0)$, $\mu_{_\perp}\equiv\mu_{_T}(|\bv|=v_0)$, $\mu_x\equiv\mu_{_T}(|\bv|=v_0)+\mu_2(|\bv|=v_0)v_0^2$, and dropped irrelevant terms arising from expanding $\lambda_1(|\bv|)$ and $\mu_{_T,2}(|\bv|)$ to higher powers of $(|\bv|-v_0)$.	
We are interested in the behavior of the state of collective motion, in which the velocity $\bv$ acquires a non-zero average value.
Without loss of generality, we will assume that the collective motion is along the $x$-direction and write the velocity field as
\beqn
\bv (\br,t)= (v_0 +u_x (\br,t)) \hat{\bx} + \bu_{_\perp} (\br,t)\,,
\label{udef}
\eeqn
 where $v_0$ is the value of $|\bv|$ at which $U(|\bv|)$ vanishes.

{\it Linear theory.}
We first study the EOM at the linear level in $\bu$.
Inserting Eq.~(\ref{udef}) into Eq.~(\ref{vEOM2}), keeping only terms linear in $\bu$, rewriting the resultant equation in Fourier space using tensor notation, and acting on it the transverse projection operator $P_{ml} = \delta_{ml} - q_mq_l/q^2$ to eliminate the isotropic pressure term, we obtain 
\beqn
-\ii (\omega- v_1q_x)u_m(\bq,\omega)=-(2a
+ \ii\lambda_4 v_0^3q_x)P_{mx}u_x(\bq, \omega)\nonumber\\ -\Gamma(\bq) u_m(\bq, \omega)+P_{ml} f_l(\bq, \omega)
\ ,\,\,\,\,\,\,\label{eq:mainu}
\eeqn
where $v_1\equiv\lambda v_0$ and
\beq
\label{eq:Gamma}
\Gamma(\bq)\equiv \mu_{_\perp}q_{_\perp}^2+\mu_xq_x^2
\ .
\eeq
The coefficient $\lambda_4$ and the longitudinal mass $a$ are defined respectively by
\beq
\lambda_4 \equiv {1\over v}\left[\dd P_2(v)\over \dd v\right]_{v=v_0}
\sep
a\equiv {v_0\over 2}\left[\dd U(v)\over \dd v\right]_{v=v_0}\,.
\eeq

\begin{figure}
	\begin{center}
		\includegraphics[scale=.55]{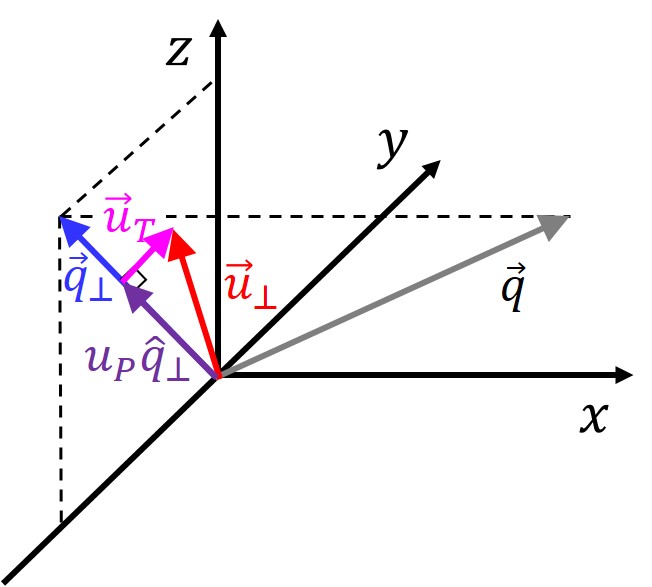}
	\end{center}
	\caption{Schematics of the vectorial decomposition discussed in this paper. Note that all vectors shown are orthogonal to $\hat{\bx}$ except $\bq$. Because soft  fluctuations must be orthogonal to both of the  direction of collective motion (parallel to $\hat{\bf x}$) and $\bq$ (the latter by the incompressibility condition), fluctuations of $\bu_{_T}$ dominate over those of other components of $\bu$.
	}
	\label{fig:vectors}
\end{figure}

To proceed we first eliminate $u_x$ in terms of the other fields using the Fourier transform of the incompressibility condition ($\nabla \cdot \bu=\nabla \cdot \bv = 0$):
\beq
u_x(\bq,t)=-\frac{\bq_{_\perp}\cdot\bu_{_\perp}(\bq,t)}{q_x}
\, .
\label{eq:ux}
\eeq
Since  we have chosen the $x$-direction to be the `stiff' direction, we expect that fluctuations of $u_x$ are small. In addition, Eq. (\ref{eq:ux}) shows that the component of $\bu_{_\perp}$ along the direction of $\bq_{_\perp}$ and $u_x$ are locked together, which suggests the fluctuations of this component of $\bu_{_\perp}$ are small as well.
To verify this, we further decompose $\bu_{_\perp}$ into components parallel and perpendicular to $\bq_{_\perp}$ (see \fig \ref{fig:vectors}):
\beqn
\bu_{_\perp}=u_{_P}\hat{\bq}_{_\perp}+\bu_{_T}
\ ,
\eeqn
where we use the subscript $P$ ($T$) to denote the component parallel (transverse) to $\hat{\bq}_{_\perp} \equiv \bq_{_\perp}/|\bq_{_\perp}|$.

We apply the projection operator
$P_{mn}^{\perp}=\delta_{mn}^{\perp}-{q_m^{\perp}q_n^{\perp}\over q_{\perp}^2}$
to Eq.~(\ref{eq:mainu}) and solve the resultant equation for $\bu_{_T}$ to obtain
\beqn
u_m^{_T}(\bq,\omega)={P_{mn}^{\perp} f_n(\bq, \omega)\over -\ii(\omega - v_1q_x)+\Gamma(\bq)}
\ ,\label{utsol}
\eeqn

Having found $\bu_{_T}$,  we now turn to $u_x$ and $u_P$. Taking the $x$ component of (\ref{eq:mainu}) and solving for $u_x$, we find
\beq
u_x(\bq,\omega)={P_{xm}(\bq)f_m(\bq,\omega)\over -\ii\left[\omega-c( \hat{\bq})q\right]+\Gamma(\bq)+2a{q_\perp^2\over q^2}}\ ,
\label{uxsol}
\eeq
where 
$c(\hat{\bq})$ is defined as
\begin{eqnarray}
c(\hat{\bq})\equiv v_1{q_x\over q}+\lambda_4v_0^3{q_{_\perp}^2q_x\over q^3}\,.
\end{eqnarray}
This, combined with the Fourier transform of the incompressibility condition, which reads $q_x u_x+q_{\perp}u_{_P}=0$, gives
\beq
u_{_P}(\bq,\omega)=-{q_x\over q_{_\perp}}{P_{xm}(\bq)f_m(\bq,\omega)\over -\ii\left[\omega - c(\hat{\bq})q\right]+\Gamma(\bq)+2a{q_{\perp}^2\over q^2}}\ .
\label{upsol}
\eeq

We can now autocorrelate these expressions (\ref{utsol},\ref{uxsol},\ref{upsol}), 
integrate the resultant correlation functions over all frequency $\omega$, and divide by $2\pi$ to get the equal time correlation functions:
\begin{eqnarray}
\langle\bu_{_T}(\bq,t)\cdot\bu_{_T}(\bq',t)\rangle
&=&{D\delta(\bq+\bq')\over \Gamma(\bq)}\,,
\label{eq:uT_fluct}
\\
\left<u_x(\bq,t)u_x(\bq',t)\right>&=&{Dq_{_\perp}^2\delta(\bq+\bq')\over \Gamma(\bq)q^2+2a q_{_\perp}^2}\,,
\label{upcorr_x}\\
\left<u_{_P}(\bq,t)u_{_P}(\bq',t)\right>&=&
{Dq_x^2\delta(\bq+\bq')\over \Gamma(\bq)q^2+2a q_{_\perp}^2}\,.
\label{upcorr_P}
\end{eqnarray}

The expressions above show that, as expected, the fluctuations of $u_{x,_P}$ are much smaller than the fluctuations of $\bu_{_T}$ for almost all directions of $\bq$
as $\bq\rightarrow\bf{0}$. In addition, the dominant field $\bu_T$ has
spatially isotropic fluctuations in this linear theory.

Now the real space fluctuations can be readily calculated:
\begin{eqnarray}
\left\la |\bu(\br, t)|^2 \right\ra
&=&{1\over (2\pi)^d}\int \dd^d q \dd^d q'\langle \bu(\bq, t)\cdot \bu(\bq', t)\rangle\nonumber\\
&\approx&{1\over (2\pi)^d}\int \dd^d q \dd^d q'\langle \bu_{_T}(\bq, t)\cdot \bu_{_T}(\bq', t)\rangle\nonumber\\
&=&{1\over (2\pi)^d}\int_{q\gtrsim \frac{1}{L}} \dd^d q\ \ \frac{D}{\Gamma(\bq)}
\ ,
\end{eqnarray}
where in the ``$\approx$" we have only kept the dominate fluctuations.
This integral converges as $L\rightarrow \infty$ for $d>2$, which implies long-range orientational order in the ordered phase in $d=3$. We will show later that this conclusion remains valid even beyond the linear theory.

{\it Nonlinear theory.}
 Since the fluctuations of $\bu_{_\perp}$ dominate over those of $u_x$, we insert Eq.~(\ref{udef}) into Eq.~(\ref{vEOM2}) and focus on the $\perp$ components of the resultant EOM. By power counting, we can show that all non-linearities arising from $U(v)$ are irrelevant, as well as those arising from the $P_2$ term; details are given in the SI. Dropping those non-linearities, and boosting to a new Galilean frame via the change of variables $x=x^\prime-v_1t$, where $v_1 = \lambda v_0$, gives
\begin{eqnarray}
\partial_{t}\bup+\lambda(\bup\cdot\nabla_{_\perp})\bup&=&-\nabla_{_\perp} P  +  \mu_{_\perp}\nabla^2_{_\perp}\bup\nonumber\\ &&+ \mu_x\partial_x^{2}\bup+\mathbf{f}_{_\perp}\,\,.
\label{upEOM2}
\end{eqnarray}
where all $x$-derivatives are now implicitly derivatives with respect to the
``pseudo-co-moving'' coordinate $x^\prime$ defined above; we have simply suppressed the primes for convenience.

We will now derive the exact scaling exponents from Eq.~(\ref{upEOM2}), using the dynamical renormalization group (DRG) \cite{FNS}.

The DRG starts by averaging the EOM over the short-wavelength fluctuations: i.e.,   those with support in the ``shell" of Fourier space $b^{-1} \Lambda \le |\vec{q}_{_\perp}| \le \Lambda$, where $\Lambda$ is an ``ultra-violet cutoff", and $b$ is an arbitrary rescaling factor. Then, one (anisotropically) rescales lengths, time,  and $\bup$ in equation (\ref{upEOM2}) according to
$\br_{_\perp}\mapsto b\br_{_\perp}$, $x\mapsto b^{\zeta}x$, $t \mapsto b^zt$ and $\bup \mapsto b^\chi \bu_{_\perp}$.
Note that, by construction, the rescaling of $\br_{_\perp}$ automatically  restores the ultra-violet cutoff  to $\Lambda$.

When evaluating the ``graphical corrections" - that is, the renormalizations that arise due to averaging over the short-wavelength fluctuations - it is extremely useful to note that
the $\lambda(\bup\cdot\nabla_{_\perp})\bup$ term
can be written as a total $\perp$ derivative:
\begin{eqnarray}
(\bu_{_\perp}\cdot\nabla_{_\perp})u^{\perp}_m
&=&\nabla^{\perp}_n\left(u^{\perp}_nu^{\perp}_m\right)- u^{\perp}_m\nabla_{_\perp}\cdot\bu_{_\perp}\nonumber\\
&=&\nabla^{\perp}_n\left(u^{\perp}_nu^{\perp}_m\right)+ u^{\perp}_m\partial_xu_x\nonumber\\
&\approx&\nabla^{\perp}_n\left(u^{\perp}_nu^{\perp}_m\right)\,\,.
\label{totder}
\end{eqnarray}
where in the second ``=" we have replaced $\nabla_{_\perp}\cdot \bu_{_\perp}$ with $-\partial_xu_x$ using the incompressibility condition, and in ``$\approx$" we have ignored $u^{\perp}_m\partial_xu_x$ since it is much smaller by power
counting than  $\nabla^{\perp}_n\left(u^{\perp}_nu^{\perp}_m\right)$, since
fluctuations of $u_{_\perp}$ dominate over those of $u_x$.

This implies that, when averaging over short-wavelength fluctuations, the $\lambda$ term can only renormalize terms that contain at least one $\perp$ spatial derivative. Since this term is the {\it only} relevant non-linear term in the model, this means that only terms that contain at least one $\perp$ spatial derivative can get any graphical renormalization {\it at all}.  In particular neither the ``$x$ viscosity" $\mu_x$, nor the noise strength $D$, can get any graphical renormalization.

Furthermore, there is no graphical correction to $\lambda$, due to the pseudo-Galileo invariance of the EOM (\ref{vEOM2}). That is, if we let
$\bup (\bx, t) \mapsto \bup (\bx, t) + \bu_0$ and simultaneously boost the
coordinate $\br \mapsto  \br-\lambda \bu_0t$, where $\bu_0$ is an arbitrary position-$\br$ independent vector in the $\perp$ plane, the EOM (\eq (\ref{upEOM2})) remains invariant. Since this symmetry of the EOM involves $\lambda$, this implies that $\lambda$ cannot be graphically renormalized.

Based on these arguments, the DRG flow equations can be written {\it exactly} as
\beqn
\label{eq:mux}
{\dd\ln\mu_x\over \dd \ell} &=& z-2\zeta\,,\\
\label{eq:mup}
{\dd\ln\mu_{_\perp}\over \dd \ell}&=&z-2+G\,,\\
\label{eq:lambda}
{\dd\ln\lambda\over  \dd \ell}&=&z-1+\chi\,,\\
\label{eq:D}
{\dd\ln D\over  \dd \ell}&=&z-2\chi-\zeta-(d-1)\,,
\eeqn
where $G$ represents graphical corrections to $\mu_{_\perp}$.
At a fixed point, the right hand sides of \eqs (\ref{eq:mux}), (\ref{eq:lambda}) and (\ref{eq:D}) must vanish. Solving the resultant simple linear equations for $z$, $\zeta$, and $\chi$ yields the exact exponents given in (\ref{exact}).

These exponents were first predicted to hold for {\it compressible} active fluids in \cite{TT1}; however, subsequent reanalysis \cite{NL} showed that the arguments for those exponents in the compressible case were not compelling, due to the presence of addition relevant non-linearities that are not total derivatives, and also violate the pseudo-galilean invariance. These exponents nonetheless appear empirically to work quite well \cite{TT2} in 2D compressible systems, and have been conjectured \cite{NL} to be exact for that case as well, but at present a compelling theoretical argument for them is lacking, in contrast to what we have presented here. A compelling argument {\it can}, and has \cite{TT5} been presented that shows that these same exponents govern the ordered phase  of ``Malthusian flocks" (i.e., active fluids with birth and death) in 2D \cite{TT5}.

With these exponents in hand we can make predictions for the scaling behavior of the velocity correlation functions
\begin{eqnarray}
C(x,\br_{\perp},t)\equiv \langle \bu_{\perp}(x',\br'_{\perp},t')\cdot\bu_{\perp}(x'',\br''_{\perp},t'')\rangle
\end{eqnarray}
where $x=x''-x'$, $\br_{\perp}=\br''_{\perp}-\br'_{\perp}$, and $t=t''-t'$.
The DRG analysis implies
\begin{eqnarray}
C(x,\br_{\perp},t)
=b^{2\chi}C(|x|b^{-\zeta},r_{\perp}b^{-1},|t|b^{-z})\,.
\end{eqnarray}
Letting $b=r_{\perp}$ in the above equation we obtain
\begin{eqnarray}
C(x,\br_{\perp},t)
=r_{\perp}^{2\chi}g\left({|x|\over r_{\perp}^\zeta},{|t|\over r_{\perp}^z}\right)
\label{Corrl1}
\end{eqnarray}
where
\begin{eqnarray}
g\left({|x|\over r_{\perp}^\zeta},{|t|\over r_{\perp}^z}\right)\equiv C\left({|x|\over r_{\perp}^\zeta},1,{|t|\over r_{\perp}^z}\right)
\end{eqnarray}
is a scaling function. The scaling behavior of $g(X,T)$ can be deduced from three limiting cases. For $r_{\perp}\to \infty$, $x=0$, and $|t|=0$, $C(|x|,r_{\perp},|t|)$ should only depend on $r_{\perp}$, which implies $g(X,T)\sim 1$. Likewise, $g(X,T)\sim X^{2\chi\over\zeta}$ for $r_{\perp}=0$, $|x|\to\infty$, and $t=0$; $g(X,T)\sim T^{2\chi\over z}$ for $r_{\perp}=0$, $x=0$, and $|t|\to\infty$. The crossover between these limiting cases can be worked out by connecting the three results of $g(X,T)$ in the parameter space of $X$ and $T$. Finally we have
\begin{eqnarray}
g(X,T)\sim\left\{
\begin{array}{ll}
1,&X\ll 1, T\ll 1\\
X^{2\chi\over\zeta},&X\gg 1, X\gg T^{1\over 2}\\
T^{2\chi\over z},&T\gg 1, T\gg X^2
\end{array}\,.
\right.\label{gScaling}
\end{eqnarray}

Plugging (\ref{gScaling}) into (\ref{Corrl1}) we find the scaling behavior of the velocity function:
\begin{eqnarray}
C(|x|,r_{\perp},|t|)\sim\left\{
\begin{array}{ll}
r_{_\perp}^{2\chi},&|x|\ll r_{_\perp}^{\zeta}, |t|\ll r_{_\perp}^z\\
|x|^{2\chi\over\zeta},&|x|\gg r_{_\perp}^{\zeta}, |x|\gg |t|^{1\over 2}\\
|t|^{2\chi\over z},&|t|\gg r_{_\perp}^z, |t|\gg |x|^2
\end{array}\,.
\right.
\end{eqnarray}
Transforming this expression back to the lab coordinates by replacing $x$ with $x-v_1t$ leads to our fundamental results (\ref{uscale}) and (\ref{exact}).

{\it Summary.}
We have studied a generic model of incompressible polar active fluids in the ordered phase, in which the continuous rotational symmetry (i.e., the rotation group SO($d$)) is broken. The resulting Goldstone modes lead to the emergence of nontrivial scaling exponents that describe the large-distance behavior of the system.
This is the first phase of an active system with complete underlying rotation invariance and anomalous scaling (that is, scaling different from that predicted by a linear theory) for which the scaling exponents have been determined exactly in $d>2$.


\section{acknowledgements}
L. C. acknowledges support by the National Science Foundation of China (under Grant No. 11474354). J.T. thanks the  Max Planck Institute for the Physics of Complex Systems  Dresden; the Department of Bioengineering at Imperial College, London; The Higgs Centre for Theoretical Physics at the University of Edinburgh; and the Lorentz Center of Leiden University, for their hospitality while this work was underway.

\onecolumngrid

\newpage
\begin{center}
	
	\textbf{\large Supplemental Materials:\\
Incompressible polar active fluids in the moving phase}
	\\
	\vspace{.1in}
	Leiming Chen\\
	{\it College of Science, China University of Mining and Technology, Xuzhou Jiangsu, 221116, P. R. China}\\
	Chiu Fan Lee\\
	{\it Department of Bioengineering, Imperial College London, South Kensington Campus, London SW7 2AZ, U.K.}\\
	John Toner\\
	{\it Department of Physics and Institute of Theoretical
		Science, University of Oregon, Eugene, OR $97403$}

\end{center}

In these supplemental materials, we present our argument that,  except for the $\lambda$ term, all of the  the non-linear terms in the equation of motion  (7) of the main text are irrelevant, in the renormalization group sense \cite{FNS}. 


Most of these  non-linearities arise from  the $U(v)\bv$ term in  (7) of the main text. It is therefore convenient to solve for  $U(v)$ directly in terms of the other fields. Following \cite{NL},
we do so by first
taking the dot product of both sides of equation  (7) of the main text with $\bv$ itself. This gives:
	\begin{eqnarray}
	{1\over 2}\left[\partial_{t}+\lambda \left(\bv\cdot\nabla\right)\right]|\bv|^2 &= &U(v)|\bv|^{2}-\bv \cdot \nabla  P-|\bv|^{2}\bv \cdot \nabla  P_2 + \mu_{_\perp}\bv\cdot\nabla_{_\perp}^{2}\bv +\mu_{_x}\bv\cdot\partial_x^{2}\bv
	+\bv\cdot\bf{f}
	\ .
	\label{v parallel elim}
	\end{eqnarray}
Using our expression
\beqn
\bv (\br,t)= (v_0 +u_x (\br,t)) \hat{\bx} + \bu_{_\perp} (\br,t)\,,
\label{udef}
\eeqn
to rewrite this in terms of the fluctuation $\bu$, dropping ``obviously irrelevant" terms - i.e., terms that differ from others in the equation of motion only by having more powers of the fields $u_x$ or $\bup$- and solving for $U(v)$, we obtain
	\begin{eqnarray}
	U(v)&=&\frac{1}{v_0}\partial_x P+\bv\cdot\nabla P_{_2}+\left({1\over  2v_0^2}\right)\left(\partial_t + v_1 \partial_x +\lambda\bu\cdot\nabla\right)\left(2v_0u_x+|\bup|^2\right) -{f_x\over v_0}\nonumber\\
	&&-\left({1\over  v_0^2}\right)\left(\mu_{_\perp}\bup\cdot\nabla_{_\perp}^{2}\bup +\mu_{_x}\bup\cdot\partial_x^{2}\bup\right)-\left({1\over  v_0}\right)\left(\mu_{_\perp}\nabla_{_\perp}^{2}u_x +\mu_{_x}\partial_x^{2}u_x\right)\,\,\,,
	\label{Usol}
	\end{eqnarray}
where $v_1=\lambda v_0$.

Inserting this back into our equation of motion (7) of the main text,  using our expression (\ref{udef}) for $\bv$, and focusing on the components of that equation perpendicular to the direction of mean motion $x$, we obtain:
	\begin{eqnarray}
	\partial_{t}\bup+v_1\partial_x\bup+\lambda(\bup\cdot\nabla_{_\perp})\bup&=&-\nabla_{_\perp} P  + \mu_{_\perp}\nabla^2_{_\perp}\bup +\mu_{_x}\partial_x^{2}\bup+\mathbf{f}_{_\perp}\,\nonumber \\&\,\,&+\bup\left[\frac{1}{v_0} \partial_x P+\left({1\over  2v_0^2}\right)\left(\partial_t + v_{_1} \partial_x +\lambda_{_1}\bup\cdot\nabla_{_\perp}\right)\left(2v_0u_x+|\bup|^2\right) -{f_x\over v_0}\right]
	\nonumber\\
	&\,\,&-\bup\left[\left({1\over  v_0^2}\right)\left(\mu_{_\perp}\bup\cdot\nabla_{_\perp}^{2}\bup +\mu_{_x}\bup\cdot\partial_x^{2}\bup\right)+\left({1\over  v_0}\right)\left(\mu_{_\perp}\nabla_{_\perp}^{2}u_x +\mu_{_x}\partial_x^{2}u_x\right)\right]\,.\nonumber\\
	\label{upEOM1}
	\end{eqnarray}

Note that the $P_2$ term has been cancelled out by a piece of the $U(v)$ term. Note also that we have replaced $(\bu\cdot\nabla)$ with $(\bup\cdot\nabla_{_\perp})$, which is justified since $u_x\ll |\bup|$ in the long-wavelength limit, as discussed in our treatment of the linear theory.

It is now straightforward to show by power counting that every term in the last two lines of Eq. (\ref{upEOM1})- that is, every term arising from the $U(v)$ and $P_2$ terms in equation (3) - is irrelevant {\it in dimensions higher than 2}.  This can be seen most easily by comparing them to various similar terms on the first line, as we will demonstrate now.

\vspace{.1in}
There are 12 terms in total:

i) $\bup\dx P$:
This term is subtle. It is tempting, but misleading, to note that this has the same number of spatial derivatives as the $\nabla_{_\perp} P$ term, but one extra power of $\bup$, and so is apparently irrelevant relative to that $\nabla_{_\perp}$ term. The subtlety is that $\nabla_{_\perp} P$ is purely parallel to $\hat{\bq}_{_\perp}$, while $\bup\dx P$ has components transverse to $\hat{\bq}_{_\perp}$,
since $\bup$ itself does. Hence, in order to prove that the $\bup\dx P$ term is truly irrelevant, we must show that it is negligible relative to some other term in (\ref{upEOM1}) that also has a transverse component. To do so, we need to obtain the power counting of $P$ itself. This can be done by taking $\nabla_{_\perp}\cdot$ both sides of (\ref{upEOM1}). Neglecting the ``obviously irrelevant" terms we solve the resultant equation for $\nabla_{_\perp}^{2}P$:
\begin{eqnarray}
\nabla_{_\perp} ^2P&=&  \left(\partial_{t}+v_1\partial_x\right)\dx u_x-\lambda\nabla_{_\perp}\cdot\left[(\bup\cdot\nabla_{_\perp})\bup\right] +\nabla_{_\perp}\cdot\mathbf{f}_{_\perp} \,\,,
\label{Psol}
\end{eqnarray}
where we have used the incompressibility condition $\nabla\cdot\bv=\nabla_{_\perp}\cdot\bup+\dx u_x=0$ to rewrite the first two terms on the right hand side in terms of $u_x$. Here in order to obtain $\nabla_{\perp}^2P$ we have also neglected $\bu_{\perp}\partial_xP/v_0$ on the right hand side of (\ref{upEOM1}), which we will show below to be irrelevant.

Inspection of this equation reveals that $\dx P$ has four terms: the first, coming from the $\partial_{t}\dx u_x$ term on the right hand side of (\ref{Psol}), power counts like $\partial_{t} u_x$. This is because $\nabla_\perp^2 P \sim \pp_t \pp_x u_x$ and derivatives in all directions power count in the same way, since scaling is isotropic according to our linear theory. Thus, this piece has the same power counting as the explicit  $\partial_{t} u_x$ term that appears later in (\ref{upEOM1}). We can thus deal with this piece of $\dx P$ at the same time as we deal with that explicit term, as we will in a few paragraphs.

Likewise, the next term on the right hand side of (\ref{Psol}), which is proportional to $\dx^2 u_x$, contributes to $\dx P$ a term which power counts like $\dx u_x$. Since an explicit term of that form appears later in (\ref{upEOM1}), we will deal with this piece of $\dx P$ at the same time as we deal with that explicit term, as we  also will in a few paragraphs.

Similarly, the $\nabla_{_\perp}\cdot\left[(\bup\cdot\nabla_{_\perp})\bup\right]$ term contributes to $\dx P$ a term which power counts exactly like the explicit
$\partial_x|\bup|^2$ term that appears later in (\ref{upEOM1}), so we can deal with it when we deal with that term in a few paragraphs.

Finally, the $\nabla_{_\perp}\cdot\mathbf{f}_{_\perp}$ term contributes to $\dx P$ a term that power
counts exactly like the 
explicit $(f_x/v_0)$ term that appears later in (\ref{upEOM1}), so we can deal with {\it it} when we deal with {\it that} term in a few paragraphs.

Now turning to the terms explicitly displayed on the last two lines of (\ref{upEOM1}):

ii) $\bup \pp_t u_x$:
This term  has the same number and type of derivatives as the $\partial_t\bup$ term on the first line, but one extra power of $u_x$, so it is irrelevant compared to that  $\partial_t\bup$ term.

iii) $\bup \pp_x u_x$:
This term has the same number of spatial derivatives as the $(\bup\cdot\nabla_{_\perp})\bup$ term on the first line, but has a $u_x$ instead of a second $\bup$. Again using the fact that  $u_x\ll |\bup|$ in the long-wavelength limit, we see that the
$\bup\dx u_x$ term is negligible relative to  the $\bup\cdot\nabla\bup$ term.

iv) $\bup (\bup \cdot \nabla_{_\perp} u_x)$:
This term differs from the $(\bup\cdot\nabla_{_\perp})\bup$ term by having one extra power of $u_x$, so it is irrelevant relative to the $(\bup\cdot\nabla_{_\perp})\bup$ term.

v) $\bup \pp_t |\bup|^2$:
This term has the same derivatives as the $\dt\bup$ term, but two more powers of $\bup$, so it is negligible relative to the $\dt\bup$ term.

vi) $\bup \pp_x |\bup|^2$:
This term has the same number of spatial derivatives as the $(\bup\cdot\nabla_{_\perp})\bup$ term but one more power of $\bup$, so it is negligible relative to the $(\bup\cdot\nabla_{_\perp})\bup$ term.

vii) $\bup\bup \cdot \nabla_\perp |\bup|^2$:
This term similarly has the same number of spatial derivatives as the $(\bup\cdot\nabla_{_\perp})\bup$ term, but now with two more powers of $\bup$, so it is even more negligible relative to the $(\bup\cdot\nabla_{_\perp})\bup$ term.

viii) $f_x\bup$:
This term 
has the same number of powers of the random force as the $\mathbf f_{\perp}$ term, but one extra power of $\bup$, so it is negligible compared to the $\mathbf f_{\perp}$ term.

ix) \& x) $\bup\left[\left(\mu_{_\perp}\bup\cdot\nabla_{_\perp}^{2}\bup +\mu_{_x}\bup\cdot\partial_x^{2}\bup\right)\right]$:
These terms have one more spatial derivative, and one more power of $\bup$, than the $(\bup\cdot\nabla_{_\perp})\bup$ term, and so are doubly negligible in comparison to that term.

xi) \& xii) $\bup\left(\mu_{_\perp}\nabla_{_\perp}^{2}u_x +\mu_{_x}\partial_x^{2}u_x\right)$:
Finally, These terms have one more spatial derivative than the $(\bup\cdot\nabla_{_\perp})\bup$ term; they also have  a $u_x$, rather than a $\bup$, and so are doubly negligible in comparison to the $(\bup\cdot\nabla_{_\perp})\bup$ term, since, as we established in the linear theory section, $u_x\ll |\bup|$ in the long-distance limit.

\vspace{.1in}
So we have, rather laboriously, established that all of the terms on the second and third lines of (\ref{upEOM1}) are irrelevant, in the RG sense, at long distances. Hence we can drop them all, leaving our effective long wavelength model for the fluctuations $\bup$ as:
\begin{eqnarray}
\partial_{t}\bup+\lambda(\bup\cdot\nabla_{_\perp})\bup&=&-\nabla_{_\perp} P  + \mu_{_\perp}\nabla^2_{_\perp}\bup+\mu_{_x}\partial_x^{2}\bup+\mathbf{f}_{_\perp}\,\,,
\label{upEOM2}
\end{eqnarray}
where we have eliminated the term $v_1\partial_x\bup$ on the right-hand side of the equality by making a Galilean transformation to a ``pseudo-co-moving'' coordinate system moving along $\hat{\bf x}$ with a constant speed $v_1$. The remainder of the analysis of this non-linear model is contained in the main text.

\end{document}